\documentclass[aps,twocolumn]{revtex4}
\usepackage{amsmath}
\usepackage{color}
\usepackage{rotating}
\usepackage{afterpage}
\usepackage{pdflscape}
\let\mc\multicolumn

\begin{document}
\title{\textit{Ab initio} effective rotational and rovibrational Hamiltonians for non-rigid systems via curvilinear second order vibrational M{\o}ller--Plesset perturbation theory}
\author{P. Bryan Changala}
\email{bryan.changala@colorado.edu}
\affiliation{JILA, National Institute of Standards and Technology and University of Colorado;\\ and Department of Physics, University of Colorado, Boulder CO 80309}
\author{Joshua H. Baraban}
\affiliation{Department of Chemistry, University of Colorado, Boulder, CO 80309}

\begin{abstract}
We present a perturbative method for \textit{ab initio} calculations of rotational and rovibrational effective Hamiltonians of both rigid and non-rigid molecules. Our approach is based on a curvilinear implementation of second order vibrational M{\o}ller-Plesset perturbation theory (VMP2) extended to include rotational effects via a second order contact transformation. Though more expensive, this approach is significantly more accurate than standard second order vibrational perturbation theory (VPT2) for systems that are poorly described to zeroth order by rectilinear normal mode harmonic oscillators. We apply this method and demonstrate its accuracy on two molecules: Si$_2$C, a quasilinear triatomic with significant bending anharmonicity, and CH$_3$NO$_2$, which contains a completely unhindered methyl rotor. In addition to these two examples, we discuss several key technical aspects of the method, including an efficient implementation of Eckart and quasi-Eckart frame embedding that does not rely on numerical finite differences.
\end{abstract}
\maketitle
\section{Introduction}

The spectroscopic perspective on a molecule is one of interacting degrees of freedom. The usual description of rovibrational motion in polyatomic molecules, in particular, consists of approximately separable anharmonic normal mode vibrations of a semi-rigid rotor. This physical picture forms the basis of the well known Watson Hamiltonian~\cite{Watson1968}, for which there are now numerous approaches for obtaining its rovibrational energies and wavefunctions. Variational methods can provide numerically exact results, but their computational expense grows exponentially with atom number, limiting converged calculations to small molecules. On the other hand, approximate methods can be significantly more efficient, while retaining sufficient accuracy for spectroscopists.

One of the most successful such approximate methods is second-order vibrational perturbation theory (VPT2)~\cite{Mills1972,Barone2005,Puzzarini2010,Barone2014}. The zeroth order Hamiltonian in VPT2 is that of a set of uncoupled harmonic oscillator vibrations and rigid top rotations. Perturbative corrections from higher order anharmonic and rovibrational terms yield anharmonic vibrational frequencies and effective rotational parameters. VPT2 is commonly used to calculate accurate \textit{ab initio} predictions of these molecular constants, which facilitate assigning and analyzing the optical and microwave spectra of polyatomic molecules. However, while the Watson Hamiltonian on which VPT2 is based is formally exact, its explicit use of rectilinear normal coordinates and single-reference Eckart embedding~\cite{Eckart1935} renders it best suited for molecules that have a well-defined equilibrium geometry from which they undergo small amplitude displacements. This leads to the observed difficulty of VPT2 with molecules that are floppy or exhibit large amplitude motion, examples of which will be presented in this work.

A more general perturbative method for solving the rovibrational problem beyond the Watson Hamiltonian is thus desirable. Non-rigid systems introduce several complications: (i) the need for curvilinear, rather than rectilinear, coordinates; (ii) a representation of the potential energy surface (PES) and nuclear kinetic energy operator (KEO) beyond a power series expansion; and (iii) a more general quasi-Eckart frame embedding not tied to a single reference geometry, an assumption that may not have a physical basis in some systems. Recent developments in curvilinear hybrid variational-perturbative methods~\cite{Pavlyuchko2015a,Pavlyuchko2015} have considered the first of these concerns. We base our approach on vibrational self-consistent field theory (VSCF) ~\cite{Bowman1978,Gerber1979,Bowman1986,Gerber1988,Carter1997,Hansen2010} augmented by second order vibrational M{\o}ller--Plesset perturbation theory (VMP2)~\cite{Norris1996,Christiansen2003,Matsunaga2002}, which is capable of addressing all three of these issues. VMP2 is analogous to the electronic MP2 method~\cite{Moller1934}: after performing a mean-field VSCF calculation, vibrational correlation effects are taken into account via Rayleigh--Schr\"{o}dinger second order perturbation theory. While originally developed for the normal mode Watson Hamiltonian, VSCF and VSCF-based methods have been successfully extended to curvilinear reaction path Hamiltonians~\cite{Carter2000c,Bowman2007}, as well as more general curvilinear vibrational Hamiltonians~\cite{Horn1989,Zuniga1991,Griffin2006,Bounouar2008,Scribano2010,Strobusch2011,Strobusch2011a,Strobusch2013}.

In this paper, we further develop the curvilinear VMP2 method to include the rotational part of the total molecular Hamiltonian. We calculate effective rotational and rovibrational Hamiltonians by applying a second order contact transformation to the rovibrational problem solved to zeroth order by VSCF. A major task for including molecular rotation is the proper choice of a body-fixed frame that sufficiently decouples rotational and vibrational motion so that the perturbative approach is viable. To this end, we implement a quasi-Eckart embedding suitable for molecules that undergo even completely unhindered large amplitude motion and therefore have no well-defined equilibrium geometry. The rovibrational KEO in this embedding, as well as the standard Eckart embedding, is evaluated numerically from an analytical method that does not rely on approximate finite difference derivatives, making it both efficient and accurate. 

After discussing the theoretical details, we present specific applications to two representative non-rigid molecules. The first is disilicon carbide, Si$_2$C. While this molecule does have a well defined Si--C--Si bent equilibrium geometry~\cite{McCarthy2015}, the barrier to linearity is relatively small ($\sim$800~cm$^{-1}$)~\cite{Reilly2015}. The low energy bending mode, which has a fundamental frequency of only 142~cm$^{-1}$~\cite{Reilly2015}, is highly anharmonic and causes problems for VPT2 in combination with a standard rectilinear quartic force field (QFF). Curvilinear VMP2 provides an improved description of both the vibrational and rotational structure of this quasilinear system. Numerically exact variational energies are readily available for this small molecule, permitting an ``apples-to-apples'' benchmark of the rovibrational VMP2 results.

The second system we discuss is nitromethane, CH$_3$NO$_2$, which exhibits nearly barrierless internal rotation of the methyl group with respect to the NO$_2$ plane~\cite{Tannenbaum1956,Jones1968,Cox1972,Rohart1975,Sorensen1983,Sorensen1983a}. As a result, standard VPT2 is qualitatively incapable of treating the large amplitude motion, total rotation, and the interactions between them. We show that the combination of a flexible choice of curvilinear internal coordinates and quasi-Eckart embedding enables VMP2 to predict an accurate effective torsional-rotational Hamiltonian for this challenging system.

\section{Theory}
We begin with a brief review of the VSCF method, which we use to generate zeroth order solutions, and the standard VMP2 perturbative correction. We then describe our extension of VMP2 to account for both vibrational correlation and rovibrational effects. This is followed by a presentation of the detailed form of the rovibrational Hamiltonian, as well as a discussion of the choice of internal coordinate systems and molecule frame embedding.

\subsection{The pure vibrational problem}
The goal of VSCF is to find the variationally optimal solution to the vibrational Hamiltonian within a Hartree product wavefunction ansatz,
\begin{align}
\Psi_0(\vec{q}\,)&= \psi_1(q_1) \psi_2(q_2) \ldots \psi_{N_m}(q_{N_m}),
\end{align}
where $\vec{q} = (q_1,\ldots,q_{N_m})$ are the $N_m$ curvilinear internal coordinates. In ket notation, we write this as
\begin{align}
\vert \Psi_0 \rangle &= \vert 1 \rangle \vert 2 \rangle \ldots \vert N_m \rangle = \prod_{k} \vert k \rangle.
\end{align}
The one-dimensional (1D) wavefunction for each vibrational degree of freedom is determined by solving an effective one-body Schr\"{o}dinger equation,
\begin{align}
\hat{h}_k \vert k \rangle = \varepsilon_k \vert k \rangle,\label{eq:hk}
\end{align}
where
\begin{align}
\hat{h}_k = \left( \prod_{l\neq k} \langle l \vert \right) H_\text{v} \left( \prod_{l\neq k} \vert l \rangle \right)
\end{align}
is the mean-field one-body Hamiltonian for mode $k$, found by computing the expectation value of the vibrational ($J=0$) Hamiltonian $H_\text{v}$ over all other degrees of freedom $l \neq k$. (The detailed form of $H_\text{v}$ is discussed below.) Because the one-body operators depend on each other's eigenfunctions, these equations are solved iteratively until self-consistency is reached. In our implementation, we solve the 1D problems using an underlying discrete variable representation (DVR) basis~\cite{Light2000}, which 
permits efficient numerical quadrature integration.

Typically, the lowest energy solution to Eq.~\ref{eq:hk} is chosen for each mode, such that the product wavefunction $\vert \Psi_0 \rangle$ corresponds to the vibrational ground state configuration. (This need not be the case, however. Excited one-body eigenfunctions can be used in the VSCF solution to generate excited configurations.) The VSCF reference configuration $\vert \Psi_0 \rangle$ together with other ``virtual'' configurations, which contain as a factor one or more excited one-body wavefunctions, form a natural orthonormal direct product basis set. Using this basis to perform a variational calculation by matrix diagonalization is termed vibrational configuration interaction (VCI)~\cite{Christoffel1982,Bowman1986,Carter1998a}. While VCI converges to the 1D-basis-set-limit vibrational wavefunctions and energies using a sufficiently large virtual configuration space, the cost of the calculation scales exponentially with the number of vibrational degrees of freedom, making converged calculations practical only for small systems.

A less expensive alternative to VCI is to correct the VSCF reference wavefunction $\vert \Psi_0 \rangle$ by second order perturbation theory. In analogy with the well known electronic structure method, this procedure is given the name vibrational second order M{\o}ller--Plesset perturbation theory or VMP2. For the pure vibrational problem, one finds the VMP2 energy by computing the standard Rayleigh--Schr\"{o}dinger perturbative corrections,
\begin{align}
E_0^\text{VMP2} = E^{(0)}_0 + E^{(2)}_0\\
E^{(0)}_0 = \langle \Psi_0 | H_\text{v} | \Psi_0 \rangle\\
E^{(2)}_0 = \sum_{\vert v \rangle \neq \vert \Psi_0 \rangle } \frac{ \vert \langle \Psi_0 \vert H_\text{v} \vert v \rangle \vert ^2}{E^{(0)}_0 - E^{(0)}_v}
\end{align}
In this formulation, there is no first order contribution because the zeroth order Hamiltonian corresponds to the diagonal of the exact Hamiltonian in the VSCF basis.

\subsection{Rovibrational contact transformation}

The standard VMP2 method described above provides approximate solutions to the vibrational ($J=0$) Hamiltonian $H_\text{v} = T_\text{v} + V$, where $T_\text{v}$ is the vibrational KEO and $V$ is the potential energy surface. The full rovibrational Hamiltonian for $J>0$, $H = H_\text{v} + T_\text{r} + T_\text{rv}$, consists of additional KEO terms accounting for rotational kinetic energy ($T_\text{r}$) and rovibrational coupling ($T_\text{rv}$). With the exception of degenerate or near-degenerate vibrational states, the energy scales of $T_\text{r}$ and $T_\text{rv}$ are typically one to several orders of magnitude smaller than the spacings between vibrational states. Thus, the rotational structure of an isolated vibrational state may be adequately predicted by accounting for interactions with other vibrational states perturbatively. This is the essential idea behind the so-called contact or Van Vleck transformation~\cite{VanVleck1929}, which forms the foundation for the development of empirical effective rotational Hamiltonians~\cite{Wilson1936,Nielsen1951,Watson1967,Watson1977,Field2011}. 

Beginning with the total rovibrational Hamiltonian,
\begin{align}
H &= H_\text{v} + T_\text{rv} + T_\text{r},
\end{align}
we define the zeroth order Hamiltonian using the VSCF reference and virtual configurations $\vert v \rangle$ as 
\begin{align}
H_0 &= \sum_v E^{(0)}_v  | v \rangle \langle v |,
\end{align}
where $E^{(0)}_v = \langle v | H_\text{v}| v \rangle$. Letting $\Delta H = H_\text{v} - H_0$, we rewrite $H$ as
\begin{align}
H &= H_0 + \Delta H + T_\text{rv} + T_\text{r}\nonumber\\
&= H_0 + \lambda H',\label{eq:H0Hp}
\end{align}
where the perturbation term $H'$ contains the vibrational correlation potential $\Delta H$ and the rovibrational and rotational KEO contributions $T_\text{rv}$ and $T_\text{r}$. We have also introduced the order sorting parameter $\lambda$, which is formally equal to 1, in Eq.~\ref{eq:H0Hp}.

\begin{figure}[htbp]
\includegraphics{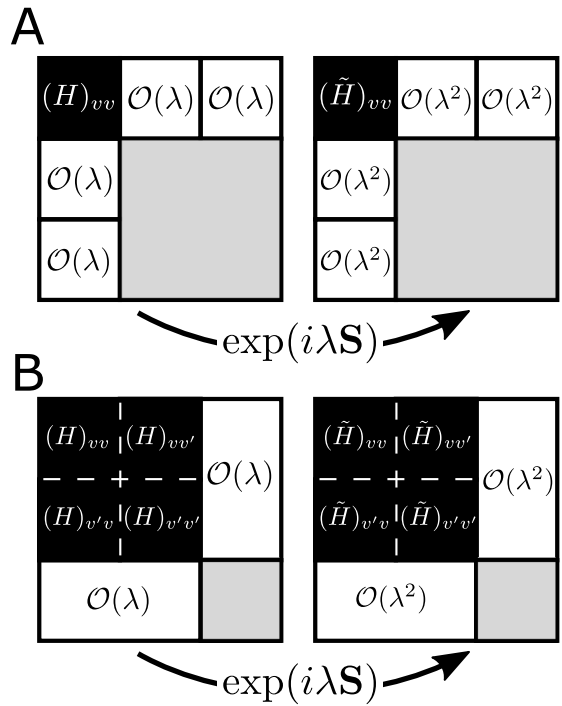}
\caption{\label{fig:heff} Block matrix form of the rovibrational Hamiltonian. (A) The case of an isolated vibrational state $\vert v \rangle$. A contact transformation by the unitary operator $\exp({i\lambda\mathbf{S}})$ eliminates the off-diagonal blocks through $\mathcal{O}(\lambda)$ and transforms the diagonal block $(H)_{vv}$ to that of the effective Hamiltonian $(\tilde{H})_{vv}$. (B) Resonant states $\vert v \rangle$ and $\vert v' \rangle$. The contact transformation reduces coupling between the resonant vibrational block and the rest of the Hamiltonian matrix. Matrix elements within the resonant vibrational block are transformed to those of the effective Hamiltonian $\tilde{H}$. }
\end{figure}
Now consider the Hamiltonian matrix representation in the zeroth order rovibrational wavefunctions $\vert v \rangle \vert r \rangle$, where $\vert r \rangle$ is some rotational basis function (usually a symmetric top wavefunction). A schematic drawing of $H$ is shown in the left of Fig.~\ref{fig:heff}A, where the vibrational block for the target vibrational state $\vert v \rangle$ is shown in solid. This block contains all matrix elements between rovibrational basis functions $\vert v \rangle \vert r \rangle$ and $\vert v \rangle \vert r' \rangle$. Matrix elements between different vibrational blocks, e.g. $\vert v \rangle \vert r \rangle$ and $\vert v ' \rangle \vert r' \rangle$, have contributions only from $H'$, which we formally assign as being of order $\mathcal{O}(\lambda)$. The contact transformation entails defining a new Hamiltonian $\tilde{H}$ that is related to the original Hamiltonian $H$ by the transformation
\begin{align}
\tilde{H} = e^{-i\lambda \mathbf{S}} H e^{i \lambda \mathbf{S}},
\end{align}
where $\mathbf{S}$ is Hermitian. As $e^{i \lambda \mathbf{S}}$ is unitary, the eigenvalues of $\tilde{H}$ are equivalent to those of $H$. The operator $\mathbf{S}$ is chosen to block diagonalize $\tilde{H}$ up to a given order in $\lambda$. In this work, we go to second order, which, as sketched in Fig.~\ref{fig:heff}, eliminates the $\mathcal{O}(\lambda)$ off-diagonal matrix elements between the target vibrational state $\vert v \rangle$ and the rest of the Hilbert space, leaving off-diagonal blocks of order $\mathcal{O}(\lambda^2)$ that are then neglected. It also introduces contributions up to $\mathcal{O}(\lambda^2)$ to the diagonal target vibrational block
\begin{align}
\tilde{H} = \tilde{H}_0 + \lambda \tilde{H}_1 +\lambda^2 \tilde{H}_2.
\end{align}

We will not derive the working equations of the Hamiltonian contact transformation here (see, for example, Appendix C of Ref.~\cite{Gordy1984}), and only state the resulting expressions for the effective Hamiltonian $\tilde{H}$. Consider first the simplest case of a single isolated vibrational state $\vert v \rangle$. The zeroth order contribution to $\tilde{H}$ is just
\begin{align}
\langle r \vert \langle v \vert \tilde{H}_0 \vert v \rangle \vert r' \rangle &=\langle r \vert\langle v \vert H_0 \vert v \rangle \vert r' \rangle = E^{(0)}_v\delta_{rr'}\label{eq:H0}.
\end{align}
The first order contribution is
\begin{align}
\langle r \vert \langle v \vert \tilde{H}_1 \vert v \rangle \vert r' \rangle &=\langle r \vert \langle v \vert H' \vert v \rangle \vert r' \rangle \nonumber\\
&=\langle r \vert  \langle v \vert \Delta H + T_\text{rv} + T_\text{r} \vert v \rangle\vert r' \rangle  \nonumber \\
&=\langle r \vert  \langle v \vert T_\text{r} \vert v \rangle\vert r' \rangle \nonumber \\
&=\langle r \vert  ( T_\text{r})_{vv} \vert r' \rangle,
\end{align}
where we have introduced the notation $(\ldots)_{vv} =  \langle v \vert \ldots \vert v \rangle$ for vibrational matrix elements. Note that $(\Delta H)_{vv} = 0$ by definition and that $(T_\text{rv})_{vv} =0$ because the vibrational factor of each term of $T_\text{rv}$ is a real, anti-Hermitian operator (see Section~\ref{sec:rovibH} below). Finally, the second order contribution is
\begin{align}
\langle r \vert  \langle v \vert  \tilde{H}_2 \vert v \rangle\vert r' \rangle &=\langle r \vert \sum_{v''\neq v} \frac{ (H')_{vv''}(H')_{v''v} }{ E_v^{(0)} - E_{v''}^{(0)}}\vert r' \rangle \label{eq:H2}.
\end{align}

As can be seen, each term of the effective Hamiltonian can be written as a pure rotational operator, independent of the actual choice of basis $\vert r \rangle$. Thus, we summarize the single-state second order effective rotational Hamiltonian for vibrational state $\vert v \rangle$ as
\begin{align}
\langle v | \tilde{H} | v \rangle = E^{(0)}_v + (T_\text{r})_{vv} + \sum_{v''\neq v} \frac{ (H')_{vv''}(H')_{v''v} }{ E_v^{(0)} - E_{v''}^{(0)}}.
\end{align}

The perturbation term $H'$ contains up to two powers of the rotational operators ($J_x$, $J_y$, $J_z$). Therefore, the second order effective Hamiltonian contains up to quartic rotational operators. By applying the known commutation relations of the body-fixed angular momentum operators, the effective rotational Hamiltonian can be reduced to six independent quadratic operators and fifteen independent quartic operators~\cite{Watson1967}. Further application of similarity transformations and commutation relations can simultaneously eliminate the three off-diagonal quadratic operators (e.g. $J_xJ_y + J_y J_x$) and ten of the quartic operators, leaving a total of three quadratic operators and five quartic operators. There are an infinite number of ways of reducing the quartic terms to produce five independent operators. Using well-known procedures~\cite{Watson1977}, we generate the effective Hamiltonian parameters corresponding to the standard A-reduced and S-reduced quartic effective rotational Hamiltonians commonly used to fit high resolution spectroscopic data.

In the case of two or more nearby or strongly interacting vibrational states, it may be necessary to include several zeroth order vibrational configurations $\{ \vert v \rangle, \vert v' \rangle, \ldots \}$ into the effective Hamiltonian. Therefore, vibrational matrix elements are required of both the diagonal type $\langle v \vert \tilde{H} \vert v \rangle$ and the off-diagonal type $\langle v \vert \tilde{H} \vert v' \rangle$. As with the single state case, the contact transformation $e^{i\lambda\mathbf{S}}$ eliminates matrix elements between the resonant states and the rest of the Hamiltonian through $\mathcal{O}(\lambda)$ and generates an effective \textit{rovibrational} Hamiltonian $\tilde{H}$ within the set of resonant vibrational states. This is illustrated schematically in Fig.~\ref{fig:heff}B. 

The effective multistate Hamiltonian is determined by a perturbation series similar to Eqs.~\ref{eq:H0}--\ref{eq:H2}. As before, the Hamiltonian can be written as effective rotational operators, independent of the choice of rotational basis functions. However, our procedure in the multistate case is slightly more complicated than the single state case. We adopt the ``diagonalize-perturb-diagonalize'' approach, wherein the resonant vibrational block of the exact vibrational Hamiltonian $H_\text{v}$ is first diagonalized to generate a new set of zeroth order eigenvectors $\{\vert w \rangle, \vert w' \rangle, \ldots\}$ that are a linear combination of the resonant VSCF configurations $\{ \vert v \rangle, \vert v' \rangle, \ldots \}$. The new zeroth order eigenvectors of course have new zeroth order energies $E^{(0)}_w = \langle w \vert H_\text{v} \vert w \rangle$ that equal the eigenvalues of the block of $H_\text{v}$ over the $\{ \vert v \rangle, \vert v' \rangle, \ldots \}$ configurations. Using the zeroth order states $\{\vert w \rangle, \vert w' \rangle, \ldots\}$, we compute the multistate effective Hamiltonian to second order as

\begin{align}
\langle w &\vert \tilde{H} \vert w' \rangle = E^0_w \delta_{ww'} + (T_\text{rv} + T_\text{r})_{w w'} \nonumber \\
&+ \frac{1}{2}\sum_{v''} \left(\frac{1}{E^{(0)}_w - E^{(0)}_{v''}} + \frac{1}{E^{(0)}_{w'} - E^{(0)}_{v''}}\right)(H')_{wv''}(H')_{v''w'},\label{eq:multiHeff}
\end{align}
where the sum over $v''$ is for vibrational VSCF configurations $\vert v'' \rangle$ not in the original resonant block $\{ \vert v \rangle, \vert v' \rangle, \ldots \}$.

In the multistate case, we cannot reduce the effective Hamiltonian without prior knowledge of the rovibrational interaction mechanisms at work. Therefore standard spectroscopic constants cannot be automatically generated from the contact transformation sums alone, as is done with a single vibrational state effective Hamiltonian. Instead, we initially proceed in a ``model agnostic'' fashion by numerically diagonalizing the resonant block of the effective Hamiltonian generated directly from Eq.~\ref{eq:multiHeff}. This diagonalization is inexpensive, the effective Hamiltonian being only of dimension $n_\text{res}(2J+1)$, where $n_\text{res}$ is the number of zeroth order vibrational states included in the resonant vibrational set. The energy eigenvalues can then be used directly or, as we do in this work, subsequently fit to a model Hamiltonian to facilitate comparison with observed spectroscopic constants obtained by fitting the same model Hamiltonian to experimental transition frequencies.

We note that by discarding the off-diagonal rovibrational matrix elements in the multistate effective Hamiltonian, the diagonal block corresponding to each vibrational level provides an estimate of the ``deperturbed'' vibrational energies and rotational constants of the zeroth order states. While these deperturbed molecular constants do not yield direct predictions of the observable rovibrational energies, they are still quite useful, often revealing patterns and physical insights into the rovibrational structure that are otherwise obscured by strong perturbations.

\subsection{Representation of the rovibrational Hamiltonian}\label{sec:rovibH}
We now consider the detailed form of the curvilinear rovibrational Hamiltonian~\cite{Podolsky1928,Lauvergnat2002,Fabri2011,Lukka1995,Watson2004}, so that we can compute the various vibrational matrix elements necessary for the VSCF calculation and VMP2 contact transformation. Let us define a general curvilinear coordinate system by specifying the Cartesian positions $\vec{x}_i$ of each atom $i = 1\ldots N$ in the body-fixed center-of-mass frame as a function of $N_m \leq 3N-6$ internal coordinates $q_k$, $k=1\ldots N_m$. We use these to define the $(N_m + 3) \times (N_m + 3)$ symmetric matrix $\mathbf{g}$ (the metric tensor) with elements given by
\begin{align}
g_{kl} &= \sum_i  m_i \partial_k \vec{x}_i \cdot \partial_l \vec{x}_i,\label{eq:gkl}\\
g_{\alpha l} &= \sum_i m_i (\hat{e}_\alpha \times \vec{x}_i) \cdot \partial_l \vec{x}_i,\label{eq:galphal}\\
g_{\alpha \beta} &= \sum_i m_i (\hat{e}_\alpha \times \vec{x}_i) \cdot  (\hat{e}_\beta \times \vec{x}_i).\label{eq:galphabeta}
\end{align}
Here and throughout, $k$, $l$, $m$, etc. are vibrational indices, having values $1\ldots N_m$, and $\alpha$, $\beta$, $\gamma$, etc. are rotational/axial indices, taking the values $x$, $y$, and $z$. The vector $\hat{e}_\alpha$ is the unit vector along the body-fixed $\alpha$-axis. $m_i$ is the mass of the $i^{th}$ atom. With knowledge of $\mathbf{g}$, we also define $g = \text{det}(\mathbf{g})$ and $\mathbf{G} = \mathbf{g}^{-1}$.

The total Hamiltonian for this coordinate system is partitioned as
\begin{align}
H = T_\text{v} + T_\text{r} + T_\text{rv} + V,
\end{align}
where the first three terms are the vibrational, rotational, and rovibrational kinetic energy operators, and $V$ is the vibrational potential energy.

We represent the vibrational kinetic energy operator $T_\text{v}$ as
\begin{align}
T_\text{v} = \frac{\hbar^2}{2} \sum_{kl} \partial_k^\dagger G_{kl} \partial_l + \frac{\hbar^2}{2}\sum_l \left(U_l\partial_l + \partial_l^\dagger U_l\right) + V_T,
\end{align}
where
\begin{align}
U_l &=-\frac{1}{4} \sum_k \left(\frac{\partial_k g}{g}\right)G_{kl},
\end{align}
and
\begin{align}
V_T &= \frac{\hbar^2}{32} \sum_{kl} \left( \frac{\partial_k g}{g}\right) \left( \frac{\partial_l g}{g} \right) G_{kl}.
\end{align}
We use $\partial_k$ to denote  $\partial / \partial q_k$, the partial derivative with respect to the internal coordinate $q_k$. The necessary derivatives of $g$ can be determined by the general algebraic relation $(\partial_k g) / g = \text{tr}(\mathbf{G} \partial_k \mathbf{g})$. The derivative matrices $\partial_k \mathbf{g}$ are themselves determined by direct differentiation of Eqs.~\ref{eq:gkl}--\ref{eq:galphabeta}. Therefore, $T_\text{v}$ as formulated here requires knowledge of both the first and second derivatives of $\vec{x}_i$ with respect to the internal coordinates. (For now, we take these as given, but this will be discussed further below.)

The rotational and rovibrational kinetic energy operators are 

\begin{align}
T_\text{r} &= \frac{-\hbar^2}{4}  \sum_{\alpha\beta} G_{\alpha\beta} \left[\frac{i\hat{J}_\alpha}{\hbar},\frac{i\hat{J}_\beta}{\hbar}\right]_+,
\end{align}
and
\begin{align}
T_\text{rv} &=\frac{-\hbar^2}{2} \sum_{k\gamma} (-\partial_k^\dagger G_{k\gamma} + G_{k\gamma}\partial_k )\frac{i\hat{J}_\gamma}{\hbar}\\
&=\frac{-\hbar}{2} \sum_{k\gamma}\lambda_{k\gamma} \frac{i\hat{J}_\gamma}{\hbar}.
\end{align}
The brackets $[\; , \;]_+$ denote an anti-commutator, while the operators $\hat{J}_\alpha$ are the body-fixed components of the total angular momentum. We have also introduced the real, anti-Hermitian vibrational operator $\lambda_{k\gamma} \equiv  (-\partial_k^\dagger G_{k\gamma} + G_{k\gamma}\partial_k )$.

Each term in the rovibrational Hamiltonian is separable into vibrational derivatives, angular momentum operators, and a single scalar function, the possibilities for the latter being (up to a constant coefficient) $G_{kl}$, $G_{k\gamma}$, $G_{\alpha \beta}$, $U_l$, $V_T$, or $V$~\cite{Strobusch2011}. Each of these quantities is a function of the $N_m$ internal coordinates $q_k$. Computing vibrational matrix elements of these functions with VSCF basis states benefits from expanding these $N_m$-dimensional functions into the so-called $n$-mode or many-body expansion~\cite{Carter1997,Strobusch2011}. For some scalar function $F(q_1,\ldots,q_{N_m})$, this expansion is
\begin{align}
F(q_1,\ldots,q_{N_a}) &= F_0 + \sum_k F_k(q_k) + \sum_{k<l} F_{kl}(q_k,q_l)\nonumber\\
&\qquad + \sum_{k<l<m} F_{klm}(q_k,q_l,q_m) + \ldots,
\end{align}
\begin{align}
F_0 &= F(q^{r}_1,\ldots,q^{r}_{N_a}),\\[1mm]
F_k &= F(q^{r}_1,\ldots,q_k,\ldots,q^{r}_{N_a}) - F_0,\\[1mm]
F_{kl} &= F(q^{r}_1,\ldots,q_k,q_l,\ldots,q^{r}_{N_a}) - F_k - F_l - F_0,\\[1mm]
F_{klm} &= F(q^{r}_1,\ldots,q_k,q_l,q_m\ldots,q^{r}_{N_a})\nonumber\\
&\qquad - F_{kl} - F_{km}  - F_{lm} - F_k - F_l - F_m - F_0,
\end{align}
where ${q_i^r}$ is a fixed internal coordinate reference geometry about which the representation is expanded. The $F_i$ functions are the ``1-body'' or first order terms, $F_{ij}$ are ``2-body'' or second order, etc. Including all possible terms up to the $N_m$-body $F_{12\ldots N_a}$ results in an exact representation of the original function. Often, only terms up to modest orders (much less than $N_m$) are necessary for an accurate, and significantly more efficient, representation of the full $N_m$-dimensional function. The accuracy of the truncated expansion depends critically on the approximate separability of the function with respect to the chosen degrees of freedom. The proper choice of internal coordinate system is therefore a primary consideration.

In some cases, a particular coordinate may be unfit for the many-body expansion, especially if a single reference value is not well determined or if no single fixed geometry respects the total symmetry of the molecule (for example, the rotation angle of an unhindered internal rotor). We generalize the many-body expansion by allowing for one or more ``dereferenced'' coordinates. In this case, for all values of the set of dereferenced coordinates, a separate many-body expansion is generated over the remaining non-dereferenced coordinates. 

\subsection{Curvilinear coordinates and transformations}
The formulation of the Hamiltonian in the previous section permits general curvilinear coordinate systems to be employed (with some exceptions in cases of accessible KEO singularities, which we do not discuss in detail here). All that is required is that the quantities $\vec{x}_i$, $\partial_k \vec{x}_i$, and $\partial_k \partial_l \vec{x}_i$ can be numerically evaluated as a function of the internal coordinates $q_k$. As a practical matter, our implementation automatically computes these quantities for standard Z-matrix coordinates (2-atom distances, 3-atom bond angles, and 4-atom dihedral angles), though routines for arbitrary coordinate systems can also be supplied.

The zeroth order VSCF calculation imposes a product ansatz on the vibrational wavefunction. Therefore, it is useful to transform a given curvilinear coordinate system to one that results in increased separability of the vibrational Hamiltonian. We now discuss two such transformations that we use in the examples below. The simplest procedure is to use a linear transformation to generate symmetrized coordinates or even curvilinear normal coordinates. These new coordinates $\vec{q}^{\,\prime}$ are related to the original coordinates $\vec{q}$ via
\begin{align}
\vec{q}^{\,\prime} = \mathbf{T} \vec{q} + \vec{t},\label{eq:Ttype}
\end{align}
where we make use of the vector notation $\vec{q} = (q_1, \ldots, q_{N_m})^T$, etc. The matrix $\mathbf{T}$ and offset vector $\vec{t}$ are both constants. In the case of transforming to curvilinear normal coordinates $\mathbf{T}$ and $\vec{t}$ are easily determined from the equilibrium geometry and the eigenvectors of the internal coordinate GF matrix~\cite{Wilson1955}. The first and second derivatives of the Cartesian positions $\vec{x}_i$ with respect to $\vec{q}\,' $ are simply to related to those with respect to $\vec{q}$ via the chain rule and Eq.~\ref{eq:Ttype}. We will refer to this linear coordinate transformation later as a ``T-type'' transformation.

A more general coordinate system is a reaction-path-like system obtained by a non-linear transformation of the original coordinates. The idea here is to categorize the original coordinates into a set of $N_\text{p}$ ``path'' and $N_\text{np}$ ``non-path'' coordinates, $\vec{q} = ( \vec{q}_\text{p},\vec{q}_\text{np})$, where $N_m = N_\text{p} + N_\text{np}$. The new set of coordinates $\vec{q}^{\,\prime}$ has the same partitioning, $\vec{q}^{\,\prime} = ( \vec{q}^{\,\prime}_\text{p},\vec{q}^{\,\prime}_\text{np})$. The path coordinates are identical in either set of coordinates, $\vec{q}_\text{p} = \vec{q}^{\,\prime}_\text{p}$, while the non-path coordinates are related via a linear transformation that itself depends on the value of the path coordinates. That is, we implicitly define $\vec{q}^{\,\prime}$ with the relation
\begin{align}
\left[ \begin{array}{c}
\vec{q}_\text{p} \\
\vec{q}_\text{np}
\end{array} \right ] &= 
\left[ \begin{array}{cc}
\mathbf{I}_{N_\text{p}} & 0\\
0 & \mathbf{L}(\vec{q}_\text{p})
\end{array}\right]
\left[ \begin{array}{c}
\vec{q}^{\,\prime}_{\text{p}}\\
\vec{q}^{\,\prime}_\text{np}
\end{array}\right]
+
\left[ \begin{array}{c}
0\\
\vec{\ell}(\vec{q}_\text{p})
\end{array}\right],\label{eq:Ltype}
\end{align}
where $\mathbf{I}_{N_\text{p}}$ is the $N_\text{p} \times N_\text{p}$ identity matrix, the transformation matrix $\mathbf{L}(\vec{q}_\text{p})$ is a function of the path coordinates, and the vector $\vec{\ell}(\vec{q}_\text{p})$ contains the reference geometry of the untransformed non-path coordinates along the (possibly multidimensional) path.

As before, the requisite derivatives of $\vec{x}_i$ with respect to these ``L-type'' coordinates can be determined in terms of the derivatives with respect to the untransformed coordinates by straightforward, if tedious, application of the chain rule with Eq.~\ref{eq:Ltype}. To do so, we require the first and second derivatives of $\mathbf{L}(\vec{q}_\text{p})$ and $\vec{\ell}(\vec{q}_\text{p})$ with respect to the path coordinates $\vec{q}_\text{p}$.

\subsection{Eckart and quasi-Eckart body-fixed frame embedding}\label{sec:eckart}

The rotational VMP2 method treats rovibrational coupling perturbatively with a zeroth order basis that separates vibrations and rotations (and in fact ignores rotations altogether). Therefore, reducing the rovibrational interaction terms present in the molecular KEO is necessary for success. A proper choice of frame  embedding should seek to minimize the rotation-rotation ($G_{\alpha \beta}$) and rotation-vibration ($G_{k\gamma}$) coupling coefficients.  In this section, we take the time to discuss this issue in some detail and, in particular, to describe our modification of the standard procedure in order to accommodate molecules without an unambiguous reference geometry.

Beginning with systems that do have a well-defined reference geometry, the preferred approach is to embed the molecule following the well-known Eckart conditions~\cite{Eckart1935}. These conditions ensure that the rotation-vibration $G_{k\gamma}$ coefficients are zero at the reference (usually equilibrium) geometry, and remain small when the molecule is slightly displaced. Eckart frame embedding is intrinsically employed by the rectilinear Watson Hamiltonian~\cite{Watson1968}. However, its implementation for general curvilinear Hamlitonians is more complicated. The problem comes down to the following: given $\vec{x}_i$ in a non-Eckart frame, find the rotation matrix $U$ that generates the Cartesian positions in the Eckart frame $\vec{x}^{\,\prime}_i= U \vec{x}_i$, where $\vec{x}^{\,\prime}_i$ satisfy the Eckart conditions. $U$ depends on $\vec{x}_i$ and therefore $\vec{q}$. In order to construct the KEO in the Eckart frame, we need the corresponding $\mathbf{g}$ matrix in this frame, and therefore the first and second derivatives of $\vec{x}^{\,\prime}_i$ with respect to the internal coordinates,
\begin{align}
\vec{x}^{\,\prime}_i&= U \vec{x}_i,\\[2mm]
\partial_k \vec{x}^{\,\prime}_i &=  \partial_k (U\vec{x}_i) \nonumber \\
&= (\partial_k U) \vec{x}_i + U(\partial_k \vec{x}_i),\\[2mm]
 \partial_m \partial_k \vec{x}^{\,\prime}_i  &= \partial_m  \partial_k (U\vec{x}_i) \nonumber \\
&= \partial_m \left[  (\partial_k U) \vec{x}_i + U(\partial_k \vec{x}_i) \right]\nonumber  \\
&= (\partial_m\partial_k U) \vec{x}_i + (\partial_k U)(\partial_m \vec{x}_i)\nonumber \\
&\qquad+ (\partial_m U)(\partial_k \vec{x}_i) + U(\partial_m \partial_k \vec{x}_i).
\end{align}
The derivatives of the non-Eckart frame positions $\vec{x}_i$ are already given. The remaining task is to determine the rotation matrix $U$ and its derivatives with respect to the internal coordinates. Several methods have appeared in the literature for determining the $U$ rotation for general coordinate systems (see, for example, Refs.~\cite{Dymarsky2005,Szalay2015,Szalay2015a,Lauvergnat2016}). We use the quaternion-based approach of Krasnoshchekov \textit{et al.}~\cite{Krasnoshchekov2014} and direct the reader to the reference for a detailed discussion of the method. Here, we only sketch the procedure for calculating $U$, as well as our extended approach to determine the first and second derivatives of $U$ analytically without the use of finite difference approximations.

In the quaternion algebra based method, $U$ is a simple function of the elements of a 4-component vector $\vec{\gamma}$. $\vec{\gamma}$ is the eigenvector with the lowest eigenvalue of the positive semi-definite real symmetric $4\times4$ matrix $C$. The matrix elements of $C$ are themselves simple functions of the non-Eckart Cartesian positions $\vec{x}_i(\vec{q})$ and the reference equilibrium geometry in the Eckart frame $\vec{x}^R_i$. This string of relations between $U$ and $\vec{x}_i(\vec{q})$ permits the derivatives of $U$ to be determined straightforwardly by repeated application of the chain rule with one exception being the derivatives of the eigenvector $\vec{\gamma}$ with respect to the internal coordinates. Fortunately, closed form expressions for the first and second derivatives of (non-degenerate) eigenvectors of real symmetric matrices are readily derived. (These are included in the appendix.) Thus, we are left with an analytical procedure, apart from a single numerical matrix diagonalization of the $4\times4$ matrix $C$, for calculating the rovibrational KEO in an Eckart frame. This approach has the benefit of avoiding the use of finite difference approximations, which can be both less accurate and less efficient.

This method for determining $U$ and its derivatives is straightforwardly generalized to allow for a non-stationary Eckart reference geometry $\vec{x}^R_i(\vec{q})$ that depends explicitly on (usually a subset of) the internal coordinates. We call this a ``quasi-Eckart'' embedding. This change from a standard Eckart embedding simply comes in when determining the matrix $C$ and its derivatives. The availability of a non-stationary $\vec{x}^R_i(\vec{q})$ is useful for molecules in which one or more coordinates exhibit large amplitude motion. As with the deferenced many-body expansions discussed earlier, this is the case for the unhindered methyl rotor in nitromethane. We discuss in more detail below our choice of a non-stationary Eckart reference geometry for this molecule.

\section{Numerical results and discussion}
The rotational VMP2 method we have described as well as the associated coordinate transformation and Eckart embedding approaches have been implemented in the program \textsc{Nitrogen}~\cite{NITROGEN}. Comparative VPT2 calculations were performed using the \textsc{CFour} progam package~\cite{CFOUR}. Atomic masses are used in all calculations.

\subsection{Disilicon carbide, Si$_2$C}
The rovibrational structure of the ground electronic state of Si$_2$C has only quite recently been experimentally characterized by optical dispersed fluorescence studies~\cite{Reilly2015}, which motivated high level electronic structure and rovibrational calculations performed partly by the authors of this work. These calculations subsequently aided the first detection of the high resolution microwave spectrum of Si$_2$C in both the laboratory~\cite{McCarthy2015} and in space~\cite{Cernicharo2015}. Due to the non-rigid, anharmonic nature of the low frequency Si--C--Si bending mode, it was found that a variational treatment of the rovibrational motion was necessary to accurately predict the rotational and vibrational energies, rather than perturbative VPT2. The deficiency of the VPT2 treatment is rooted in the rectilinear coordinates and quartic force field (QFF) representation of the potential energy surface. Here we show that rotational VMP2, in light of its improved zero order picture relative to VPT2, can provide accurate rotational and vibrational predictions, while being significantly less expensive than variational rovibrational calculations.

The primitive Si$_2$C coordinate system used in the VMP2 calculations consists of the two $r_\text{SiC}$ bond lengths and the $\angle$Si--C--Si angle. The curvilinear normal coordinates are constructed by a T-type transformation of the primitive coordinates, with the transformation matrix determined by a GF normal mode calculation at the bent Si$_2$C equilibrium geometry. We use a standard Eckart embedding referenced to the Si$_2$C equilibrium geometry in its principal axis system. All many-body expansions were taken to third order (i.e. exact), so that no approximations are made in the construction of the rovibrational Hamiltonian. The size of the 1D DVR basis sets in the VSCF portion of the calculation as well as the size of the virtual configuration space in the VMP2 sums were both enlarged until the vibrational energies converged to the reported precision. The VMP2, VPT2, and variational calculations were all performed with a potential surface calculated at the frozen core (FC)-CCSD(T)/cc-pVQZ level of theory, with the variational results taken from our previous work~\cite{McCarthy2015,Reilly2015}. 

Table~\ref{tab:si2c} compares the quartic rotational Hamiltonian predictions for the vibrational ground state as well as the excited fundamental levels. The VMP2 predictions are in excellent agreement with the corresponding benchmark variational values: the vibrational fundamentals differ by less than 0.03 cm$^{-1}$, and the $A$, $B$, and $C$ constants differ by only $\sim$0.01--0.1\% for both the ground and fundamental vibrational states. In contrast, the VPT2/QFF predictions are considerably less accurate. The vibrational fundamentals have errors on the order of $\sim$10~cm$^{-1}$. Most strikingly, the predicted ground state $A$ constant has an error of several hundreds of~MHz, and the predicted $A$ constant for the $\nu_2$ bending level is more than 2~GHz low. These difficulties arise from the quasilinearity of Si$_2$C, characterized by the small energy difference of about 800 cm$^{-1}$ between the linear saddle point and the bent equilibrium geometry. As the Si--C--Si angle approaches 180$^\circ$, the $a$ axis moment of inertia vanishes. Thus, the $A$ constant is particularly sensitive to the vibrational wavefunction, so much so that even the zero-point bending motion in the ground vibrational state is not well described by VPT2 relative to VMP2 and the variational benchmark. 

The centrifugal distortion (CD) constants provide an additional measure of the accuracy of the rotational VMP2 treatment. The error relative to the the benchmark value for the $D_K$ constant, which was the only CD parameter well determined in the fits of the variational energies, is small for each of the ground and fundamental vibrational states. Furthermore, the agreement of the VMP2 predictions with the observed values for all five quartic CD constants of the ground state is quite good.  It is difficult to compare the VMP2 CD constants with the observed values for fundamental vibrational levels as the current set of observed microwave transitions in these states has permitted only a partial determination of the quartic CD parameters. Given the accuracy of the VMP2 ground state predictions, the excited state predictions could possibly aid in the experimental search for additional transitions.

\begin{turnpage}
\begin{table*}[htpb]
\caption{\label{tab:si2c} Effective rotational Hamiltonian parameters (I$^r$, S-reduction) and vibrational energies for the ground and fundamental vibrational levels of Si$_2$C. All calculated values are at the FC-CCSD(T)/cc-pVQZ level of theory. Only the $D_K$ centrifugal distortion (CD) constant was well-determined by fits to the variational energies, so the remaining four CD constants are not reported. Vibrational energies are given in cm$^{-1}$, and all rotational parameters are given in MHz. }
\resizebox{9.2in}{!}{
\begin{tabular}{ll|rrrr|rrrr|rrrr|rrrr}
\hline
 \mc2c{}& \mc4c{VMP2} & \mc4c{Variational} & \mc4c{VPT2 (w/ QFF)} & \mc4c{Observed} \\
\hline
& & \mc1c{$v=0$} & \mc1c{$\nu_2$} & \mc1c{$\nu_1$} &\mc1{c|}{$\nu_3$}& \mc1c{$v=0$} & \mc1c{$\nu_2$} & \mc1c{$\nu_1$} &\mc1{c|}{$\nu_3$}& \mc1c{$v=0$} & \mc1c{$\nu_2$} & \mc1c{$\nu_1$} &\mc1{c|}{$\nu_3$}& \mc1c{$v=0^{a}$} & \mc1c{$\nu_2^{b,c}$} & \mc1c{$\nu_1^{b,c}$} &\mc1c{$\nu_3^{c}$}\\
$\nu_0$ &(cm$^{-1}$) && 140.482 & 828.286 & 1198.168 & & 140.494 & 828.243 & 1198.141 & & 148.821 & 846.623 & 1213.935 && 140(2) & 830(2) & ---\\
$A$ &(MHz) & 63613.974 & 70626.586 & 67996.200 & 59567.693 & 63627.255 & 70667.629 & 68099.948 & 59584.440 & 63313.287 &68435.877&66238.769&59539.742& 64074.3366(44) & 71230.1156(259)& 68646.973(59) & 59990.8638(212) \\
$B$ && 4338.530 & 4269.018 & 4239.314 & 4407.445 & 4338.654 & 4270.398 & 4239.440 & 4407.107 & 4339.899 &4276.984&4253.499&4407.524& 4395.51772(41) & 4323.2963(34) & 4291.6565(85) & 4465.9029 (45)\\
$C$ && 4050.610 & 3996.181 & 3974.878 & 4095.133 & 4050.886 & 3997.876 & 3975.396 & 4095.458 & 4051.837 &4002.498&3985.280&4095.394& 4102.130980(62) & 4045.78422(259) & 4022.82959(260) & 4147.7873 (35)\\
$A_v -A_0$ &&&7012.613 & 4382.227 & -4046.280 &  & 7040.374 &4472.693 &-4042.816 & & 5122.590 & 2925.482 & -3773.545 && 7155.7790 & 4572.6364 & -4083.4728\\
$B_v -B_0$&&&-69.512 & -99.216 & 68.915 & & -68.257 & -99.214 & 68.452  & & -62.915 & -86.400 & 67.625  & & -72.2214 & -103.86122 & 70.3852\\
$C_v - C_0$&&&-54.429 & -75.731 & 44.524 & & -53.010 & -75.490 & 44.572 & & -49.339 & -66.557 & 43.557 & & -56.34676 & -79.30139 & 45.65632\\
$D_J$ &$\times 10^3$ & 9.088 & 9.521 & 9.840 & 8.640 & & & & & 8.869 & & & & 9.66776(224) &11.185(53)&11.528(187)&[9.83727]\\
$D_{JK}$& & -0.812 & -1.180 & -1.069 & -0.649 &&&&& -0.685&&&& -0.856833(73) &-1.27509(97)&-1.1834(35) &[-0.85328]\\
$D_K$ && 22.713 & 46.335 & 35.842 & 15.505 & 21.649 & 42.026 & 34.628 & 15.500 &16.189&&&& 23.58788(178) &[43.7]&[23.58788] &[15.9]\\
$d_1$&$\times 10^3 $& -1.449 & -1.545 & -1.538 & -1.374 &&&&& -1.403 &&& & -1.52630(34) &[-1.53007]&[-1.52630]&[-1.51220]\\
$d_2$&$\times10^5$& -2.0 & -1.5 & -2.1 & -2.0 &&&&& -2.27&&&& -3.18(32) &[-3.4169]&[-3.18]&[-2.34]\\
\hline
\multicolumn{18}{l}{$^a$ Ground state rotational constants from Refs.~\cite{McCarthy2015,Cernicharo2015}}\\
\multicolumn{18}{l}{$^b$ Vibrational energies from Ref.~\cite{Reilly2015}}\\
\multicolumn{18}{l}{$^c$ Excited state rotational constants from Ref.~\cite{McCarthyComm}. The currently limited number of observed transitions prevents a fit of all five quartic centrifugal distortion constants. }\\
\multicolumn{18}{l}{\hphantom{$^c$ } Values in ``[\ldots]'' brackets were held constrained during the fit.}
\end{tabular}}
\end{table*}
\end{turnpage}

It is worth noting that the magnitude of the error between the VMP2 and variational vibrational frequencies and rotational constants is typically far less than that with respect to the observed values. That is, the rotational VMP2 method, for this molecule at least, is presumably accurate enough such that the underlying potential energy surface/quantum chemistry remains the major source of \textit{ab initio} error. Thus, even more accurate predictions can be estimated by applying corrections based on higher level electronic structure calculations. For example, it is an established practice~\cite{Puzzarini2008} to combine zero-point corrections to rotational constants calculated at a lower level of theory with equilibrium rotational constants calculated at a higher level. Table~\ref{tab:b0corr} summarizes the results of this procedure for the ground state rotational constants of Si$_2$C. Here, the zero-point corrections determined by VPT2 and VMP2 at the FC-CCSD(T)/cc-pVQZ level of theory are combined with equilibrium rotational constants from the CCSD(T)/cc-pwCV5Z + $\Delta$Q/cc-pVTZ best estimate structure of Ref.~\cite{McCarthy2015}. While both VMP2 and VPT2 perform comparably for the $B$ and $C$ constants (errors of magnitude $\sim$$0.1\%$; presumably limited by the level of theory used for the PES), VMP2 provides a much better prediction than VPT2 for the $A$ constant for the same reasons discussed above.

\begin{table}[htpb]
\caption{\label{tab:b0corr} Best estimate \textit{ab initio} Si$_2$C ground state rotational constants via VMP2 and VPT2. These values combine FC-CCSD(T)/cc-pVQZ zero-point corrections with CCSD(T)/cc-pwCV5Z + $\Delta$Q/cc-pVTZ equilibrium structures~\cite{McCarthy2015}. All values are given in MHz. The fractional errors relative to the observed values (Table~\ref{tab:si2c}) are shown in parentheses.}
\begin{tabular}{c|rr|rr}
Parameter& \mc2{c|}{VMP2} & \mc2c{VPT2}\\
\hline
$A$ & 64061.516 &($-0.02$\%) & 63760.830 &($-0.49$\%) \\
$B$ & 4389.722 &($-0.13$\%)& 4391.091 &($-0.10$\%)\\
$C$ & 4097.182 &($-0.12$\%)& 4098.410 &($-0.09$\%)\\
\hline
\end{tabular}
\end{table}

\subsection{Nitromethane, CH$_3$NO$_2$}

We now turn to nitromethane, CH$_3$NO$_2$, an even more difficult challenge for \textit{ab initio} rovibrational calculations. Rotation of the CH$_3$ group with respect to the ONO plane encounters a remarkably small barrier of only 2.1~cm$^{-1}$~\cite{Sorensen1983}. The effective rotational constant of the internal rotation, approximately the inverse of the moment of inertia of the methyl group, is roughly 5.6~cm$^{-1}$, which puts the rotor squarely in the unhindered regime. The resulting torsion-rotation-vibration structure of the molecule is thus quite different than that of a semi-rigid system and makes for an interesting and important test case for rotational VMP2.

\begin{figure}[htpb]
\includegraphics[width=3.3in]{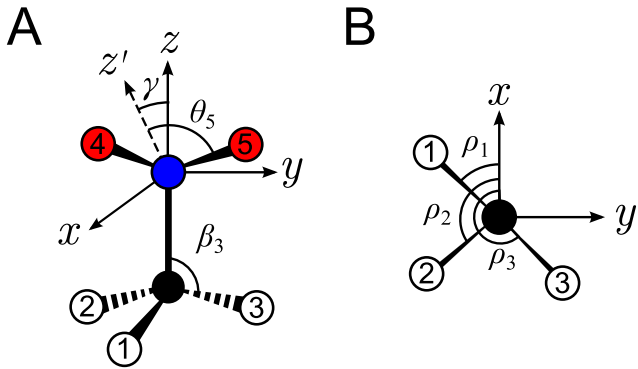}
\caption{\label{fig:ch3no2} Angular variables of the primitive coordinate system for CH$_3$NO$_2$. (A) $\gamma$ is the angle of the $z'y$ plane with respect to the $zy$ plane, rotated in a right hand sense about $+y$. The oxygen atoms, labeled \textbf{4} and \textbf{5}, lie in the $z'y$ plane and make angles $\theta_4$ (not shown) and $\theta_5$, respectively, with the $z'$ axis. The three hydrogen atoms are labeled \textbf{1}, \textbf{2}, and \textbf{3}. $\beta_i$ equals the NCH$_i$ angle ($\beta_1$, $\beta_2$ not shown). (B) Looking from the $-z$ direction, the hydrogen atoms are rotated counter-clockwise about the $z$ axis by angles $\rho_1$, $\rho_2$, and $\rho_3$, with $\rho_i = 0$ lying in the $xz$ plane.}
\end{figure}

The primitive coordinate system we use for CH$_3$NO$_2$ includes the six valence bond lengths, as well as the nine angular coordinates defined in Fig.~\ref{fig:ch3no2}. These coordinates are readily defined in a standard Z-matrix, albeit with appropriate use of dummy atoms. The hydrogen dihedral coordinates $\rho_i$ are symmetrized as $\rho_s = (\rho_1 + \rho_2 + \rho_3 - 2\pi) /3$, $\rho_a = (\rho_2 - \rho_3 + 2\pi/3)/\sqrt{2}$, and $\rho_b = -(2\rho_1 - \rho_2 - \rho_3 + 2\pi) / \sqrt{6}$. $\rho_s$ corresponds to the large amplitude internal rotation torsion angle, while $\rho_a$ and $\rho_b$ describe small amplitude HCH bending. An L-type transformation is used to generate fourteen orthogonal curvilinear normal mode coordinates as a function of the path coordinate $\rho_s$. Quasi-Eckart embedding was employed with a reference geometry that depended only on the $\rho_s$ coordinate. In particular, $\vec{x}_i^R(\rho_s)$ corresponded to the principal axis system (PAS) positions at the given value of $\rho_s$ with all other coordinates fixed at their relaxed values averaged over $\rho_s \in [0,2\pi]$. Note that because of the $C_{3v}$ symmetry of CH$_3$ group and the $C_{2v}$ CNO$_2$ frame, the reference positions of the heavy atoms do not change as a function of $\rho_s$. The many-body expansions of the Hamiltonians were expanded up to 4-body terms for $V$, 3-body terms for the diagonal $G_{kk}$ matrix elements and all of the rovibrational $G_{k\alpha}$ and rotational $G_{\alpha \beta}$ matrix elements, and 2-body terms for the off-diagonal $G_{kl}$ matrix elements and the $U_l$ and $V_T$ functions. The $\rho_s$ coordinate was dereferenced in these expansions, as defined in Section~\ref{sec:rovibH}, so that each $n$-body expansion contained terms that are each a function of $\rho_s$ and $(n-1)$-body expansions in the non-deferenced coordinates.

The torsion-vibration ground state was used as the reference VSCF wavefunction. Excited torsional states were also included into the resonant vibrational block for the VMP2 calculation to accommodate the strong interactions between the torsional and rotational degrees of freedom. The resulting multistate torsion-rotation effective Hamiltonian was diagonalized for $J = 0-5$ to generate torsion-rotation energy levels, which were then fit to empirical effective Hamiltonians~\cite{Rohart1975,Sorensen1983} to compare with measured spectroscopic constants. We cannot compare the VMP2 results to benchmark variational calculations, as the relatively large size of nitromethane (7 atoms) makes converged full dimensional rovibrational calculations prohibitively expensive. We note, however, that 14-dimensional vibrational ($J=0$) calculations, which excluded the internal rotation motion, have recently been reported~\cite{Wang2015}.

Our rotational VMP2 results, calculated with the CCSD(T)-F12b/HaDZ surface of Wang, Carter, and Bowman~\cite{Wang2015}, are summarized for the ground vibrational state in Table~\ref{tab:ch3no2}. We also include standard VPT2 calculations at the FC-CCSD(T) level of theory using the ANO1 basis set~\cite{Almlof1987a,McCaslin2013}, along with measured constants determined by microwave spectroscopy~\cite{Rohart1975,Sorensen1983}. The constants of the A-reduced Hamiltonian in Table~\ref{tab:ch3no2} have there usual meaning~\cite{Watson1977}, with the addition of $F$, the effective rotational constant of the methyl rotor, and $A'$, which describes the first order Coriolis coupling between the methyl rotor and total molecular rotation~\cite{Rohart1975}.

Without benchmark variational results, we cannot separate the error associated with the rotational VMP2 method from that of the underlying PES surface. However, the uniformly very good agreement of the VMP2 results with the observed values indicates that the VMP2 method is probably quite accurate for even this pathologically non-rigid molecule. In contrast, VPT2 has significant shortcomings here, as can be seen most clearly by the rotational constants involving angular momentum about the top axis ($a$ axis): $A$, $\Delta_{JK}$ and $\Delta_{K}$. These unphysical predictions can be directly linked to the fact that the VPT2 zeroth order model relies on a well defined equilibrium geometry and small amplitude rectilinear motions. Both of these approximations fail completely for CH$_3$NO$_2$ and its internal rotation. On the other hand, the zeroth order approximations of the rotational VMP2 method are only that the vibrational motions, in the given coordinate system, are separable. This turns out to be quite a good model for CH$_3$NO$_2$ with our choice of torsional coordinate $\rho_s$ and the orthogonal curvilinear normal coordinates generated by the L-type transformation.

The rotational VMP2 method is directly applicable to excited vibrational levels of CH$_3$NO$_2$, which have been the subject of recent high resolution spectroscopic~\cite{Dawadi2015,Spaun2016} and theoretical studies~\cite{Wang2015}. Our ongoing work on these states is beyond the scope of this paper, and we will discuss them in a future publication. Of particular interest are the degenerate vibrational fundamentals (of which there are three: a degenerate CH stretch, degenerate HCH bend, and degenerate CH$_3$ rock). In these states, vibrational, torsional, and rotational angular momentum are all coupled, creating complicated and interesting spectroscopic patterns. The use of resonant multistate rotational VMP2 is a promising tool to explore these types of nuclear motion.

\begin{table}[htpb]
\caption{\label{tab:ch3no2} Effective Hamiltonian parameters (I$^r$, A-reduction) of the ground torsion-vibration state of CH$_3$NO$_2$. VMP2 was performed on the CCSD(T)-F12b/HaDZ surface of Wang, Carter, and Bowman~\cite{Wang2015}, while VPT2 was performed at the FC-CCSD(T)/ANO1 level of theory. Unphysical VPT2 predictions are indicated with an asterisk (*).}
\begin{tabular}{llrrr}
\hline
&&\mc1c{VMP2} & \mc1c{VPT2} & \mc1c{Observed}\\
\hline\\[-4mm]
$A$& (MHz) & 13330.199$^\dagger$ & 12189.759* & 13341.881(30)$^a$\\
$B$ &&10507.162$^\dagger$  & 10464.166 & 10544.375(30)$^a$ \\
$C$ &&5862.303$^\dagger$  & 5848.808 & 5875.76(29)$^a$\\
$F$ && 166895.9$^\ddagger$ & --- & 166703.3(17)$^b$\\
$A'$ && 13248.90$^\ddagger$ & --- & 13283.03(75)$^a$\\
$\Delta_J$ &(kHz)& 5.497$^\dagger$  & 5.666 & 6.14(62)$^a$\\
$\Delta_{JK}$& & 17.840$^\dagger$  & 952.725* & 17.75(37)$^a$\\
$\Delta_K$ && -10.656$^\dagger$  & -949.388* & -7.54(40)$^a$\\
$\delta_J$ && 2.243$^\dagger$  & 2.296 & 2.467(44)$^a$\\
$\delta_K$& & 15.765$^\dagger$  & -268.953* & 15.75(19)$^a$\\
\hline
\mc5l{$^\dagger$ Computed using only $m=0$, $J=0-5$ states}\\
\mc5l{$^\ddagger$ Computed using $m = 0, \pm1$, $J=0-5$ states.}\\
\mc5l{$^a$ Ref.~\cite{Rohart1975}}\\
\mc5l{$^b$ Ref.~\cite{Sorensen1983}}
\end{tabular}
\end{table}

\section{Conclusions}
We have described an extension of curvilinear VMP2 that generates effective rotational and rovibrational Hamiltonians for non-rigid systems. The accuracy of this approach has been demonstrated for two benchmark molecules, Si$_2$C and CH$_3$NO$_2$. The primary expense of the calculation remains that of calculating the adiabatic PES, and the effect of the many-body expansion order of the PES on the accuracy of the effective Hamiltonians is worth further investigation.

An important consideration with curvilinear VMP2 is the choice of internal coordinate system, which is perhaps the most significant factor in determining the accuracy of the zeroth order solution. While this precludes this approach from being a ``black-box'' method, it offers a substantial degree of flexibility, and we envision rotational VMP2 to be generally applicable to a variety of non-rigid systems that are otherwise inappropriate for perturbative rovibrational calculations or too large for direct variational treatment.

\section{Acknowledgements}
P.B.C. is supported by the NSF GRFP (award no. DGE1144083). The authors would like to thank Michael McCarthy for making available unpublished results on Si$_2$C, as well as Xiaohong Wang and Joel Bowman for providing their CH$_3$NO$_2$ potential energy surface.


\begin{thebibliography}{10}

\bibitem{Watson1968}
J.~K. Watson.
\newblock {Simplification of the molecular vibration-rotation Hamiltonian}.
\newblock \emph{Mol. Phys.} \textbf{15}, 479--490 (1968).

\bibitem{Mills1972}
I.~M. Mills.
\newblock {{Vibration-Rotation Structure in Asymmetric- and Symmetric-Top
  Molecules}}.
\newblock In \emph{{Molecular {Spectroscopy}: {Modern} {Research}}}, edited by
  K.~N. Rao and C.~W. Mathews, chapter 3.2, pages 115--140 (Academic Press, New
  York,\,, 1972).

\bibitem{Barone2005}
V.~Barone.
\newblock {Anharmonic vibrational properties by a fully automated second-order
  perturbative approach}.
\newblock \emph{J. Chem. Phys.} \textbf{122}, 014108 (2005).

\bibitem{Puzzarini2010}
C.~Puzzarini, J.~F. Stanton, and J.~Gauss.
\newblock {Quantum-chemical calculation of spectroscopic parameters for
  rotational spectroscopy}.
\newblock \emph{Int. Rev. Phys. Chem.} \textbf{29}, 273--367 (2010).

\bibitem{Barone2014}
V.~Barone, M.~Biczysko, and J.~Bloino.
\newblock {Fully anharmonic IR and Raman spectra of medium-size molecular
  systems: accuracy and interpretation.}
\newblock \emph{Phys. Chem. Chem. Phys.} \textbf{16}, 1759--87 (2014).

\bibitem{Eckart1935}
C.~Eckart.
\newblock {Some Studies Concerning Rotating Axes and Polyatomic Molecules}.
\newblock \emph{Phys. Rev.} \textbf{47}, 552--558 (1935).

\bibitem{Pavlyuchko2015a}
A.~I. Pavlyuchko, S.~N. Yurchenko, and J.~Tennyson.
\newblock {Hybrid variational-perturbation method for calculating
  ro-vibrational energy levels of polyatomic molecules}.
\newblock \emph{Mol. Phys.} \textbf{113}, 1559--1575 (2015).

\bibitem{Pavlyuchko2015}
A.~I. Pavlyuchko, S.~N. Yurchenko, and J.~Tennyson.
\newblock {A hybrid variational-perturbation calculation of the ro-vibrational
  spectrum of nitric acid.}
\newblock \emph{J. Chem. Phys.} \textbf{142}, 094309 (2015).

\bibitem{Bowman1978}
J.~M. Bowman.
\newblock {Self-consistent field energies and wavefunctions for coupled
  oscillators}.
\newblock \emph{J. Chem. Phys.} \textbf{68}, 608--610 (1978).

\bibitem{Gerber1979}
R.~Gerber and M.~Ratner.
\newblock {A semiclassical self-consistent field (SC SCF) approximation for
  eigenvalues of coupled-vibration systems}.
\newblock \emph{Chem. Phys. Lett.} \textbf{68}, 195--198 (1979).

\bibitem{Bowman1986}
J.~M. Bowman.
\newblock {The self-consistent-field approach to polyatomic vibrations}.
\newblock \emph{Acc. Chem. Res.} \textbf{19}, 202--208 (1986).

\bibitem{Gerber1988}
R.~B. Gerber and M.~A. Ratner.
\newblock {Self-Consistent-Field Methods for Vibrational Excitations in
  Polyatomic Systems}.
\newblock \emph{Adv. Chem. Phys.} \textbf{70}, 97--132 (1988).

\bibitem{Carter1997}
S.~Carter, S.~J. Culik, and J.~M. Bowman.
\newblock {Vibrational self-consistent field method for many-mode systems: A
  new approach and application to the vibrations of CO adsorbed on Cu(100)}.
\newblock \emph{J. Chem. Phys.} \textbf{107}, 10458--10469 (1997).

\bibitem{Hansen2010}
M.~B. Hansen, M.~Sparta, P.~Seidler, D.~Toffoli, and O.~Christiansen.
\newblock {New Formulation and Implementation of Vibrational Self-Consistent
  Field Theory.}
\newblock \emph{J. Chem. Theory Comput.} \textbf{6}, 235--248 (2010).

\bibitem{Norris1996}
L.~S. Norris, M.~A. Ratner, A.~E. Roitberg, and R.~B. Gerber.
\newblock {M{\o}ller--Plesset perturbation theory applied to vibrational
  problems}.
\newblock \emph{J. Chem. Phys.} \textbf{105}, 11261--11267 (1996).

\bibitem{Christiansen2003}
O.~Christiansen.
\newblock {M{\o}ller--Plesset perturbation theory for vibrational wave
  functions}.
\newblock \emph{J. Chem. Phys.} \textbf{119}, 5773--5781 (2003).

\bibitem{Matsunaga2002}
N.~Matsunaga, G.~M. Chaban, and R.~B. Gerber.
\newblock {Degenerate perturbation theory corrections for the vibrational
  self-consistent field approximation: Method and applications}.
\newblock \emph{J. Chem. Phys.} \textbf{117}, 3541--3547 (2002).

\bibitem{Moller1934}
C.~M{\o}ller and M.~S. Plesset.
\newblock {Note on an Approximation Treatment for Many-Electron Systems}.
\newblock \emph{Phys. Rev.} \textbf{46}, 618--622 (1934).

\bibitem{Carter2000c}
S.~Carter and N.~C. Handy.
\newblock {The vibrations of H$_2$O$_2$, studied by ``multimode,'' with a large
  amplitude motion}.
\newblock \emph{J. Chem. Phys.} \textbf{113}, 987--993 (2000).

\bibitem{Bowman2007}
J.~M. Bowman, X.~Huang, N.~C. Handy, and S.~Carter.
\newblock {Vibrational Levels of Methanol Calculated by the Reaction Path
  Version of MULTIMODE, Using an ab initio, Full-Dimensional Potential}.
\newblock \emph{J. Phys. Chem. A} \textbf{111}, 7317--7321 (2007).

\bibitem{Horn1989}
T.~R. Horn, R.~B. Gerber, and M.~A. Ratner.
\newblock {Vibrational states of very floppy clusters: Approximate separability
  and the choice of good curvilinear coordinates for XeHe$_2$, I$_2$He}.
\newblock \emph{J. Chem. Phys.} \textbf{91}, 1813--1823 (1989).

\bibitem{Zuniga1991}
J.~Z{\'u}{\~{n}}iga, A.~Bastida, A.~Requena, and A.~Hidalgo.
\newblock {Self-consistent-field calculation of vibrational bound states for
  triatomic molecules using transformed Jacobi coordinates}.
\newblock \emph{J. Phys. Chem.} \textbf{95}, 2292--2297 (1991).

\bibitem{Griffin2006}
C.~D. Griffin, R.~Acevedo, D.~W. Massey, J.~L. Kinsey, and B.~R. Johnson.
\newblock {Multimode wavelet basis calculations via the molecular
  self-consistent-field plus configuration-interaction method}.
\newblock \emph{J. Chem. Phys.} \textbf{124}, 134105 (2006).

\bibitem{Bounouar2008}
M.~Bounouar and C.~Scheurer.
\newblock {The impact of approximate VSCF schemes and curvilinear coordinates
  on the anharmonic vibrational frequencies of formamide and thioformamide}.
\newblock \emph{Chem. Phys.} \textbf{347}, 194--207 (2008).

\bibitem{Scribano2010}
Y.~Scribano, D.~M. Lauvergnat, and D.~M. Benoit.
\newblock {Fast vibrational configuration interaction using generalized
  curvilinear coordinates and self-consistent basis.}
\newblock \emph{J. Chem. Phys.} \textbf{133}, 094103 (2010).

\bibitem{Strobusch2011}
D.~Strobusch and C.~Scheurer.
\newblock {Hierarchical expansion of the kinetic energy operator in curvilinear
  coordinates for the vibrational self-consistent field method}.
\newblock \emph{J. Chem. Phys.} \textbf{135}, 124102 (2011).

\bibitem{Strobusch2011a}
D.~Strobusch and C.~Scheurer.
\newblock {The hierarchical expansion of the kinetic energy operator in
  curvilinear coordinates extended to the vibrational configuration interaction
  method.}
\newblock \emph{J. Chem. Phys.} \textbf{135}, 144101 (2011).

\bibitem{Strobusch2013}
D.~Strobusch, M.~Nest, and C.~Scheurer.
\newblock {The adaptive hierarchical expansion of the kinetic energy operator.}
\newblock \emph{J. Comput. Chem.} \textbf{34}, 1210--1217 (2013).

\bibitem{McCarthy2015}
M.~C. McCarthy, J.~H. Baraban, P.~B. Changala, J.~F. Stanton, M.-A.
  Martin-Drumel, S.~Thorwirth, C.~A. Gottlieb, and N.~J. Reilly.
\newblock {Discovery of a Missing Link: Detection and Structure of the Elusive
  Disilicon Carbide Cluster}.
\newblock \emph{J. Phys. Chem. Lett.} \textbf{6}, 2107--2111 (2015).

\bibitem{Reilly2015}
N.~J. Reilly, P.~B. Changala, J.~H. Baraban, D.~L. Kokkin, J.~F. Stanton, and
  M.~C. McCarthy.
\newblock {The ground electronic state of Si$_2$C: Rovibrational level
  structure, quantum monodromy, and astrophysical implications}.
\newblock \emph{J. Chem. Phys.} \textbf{142}, 231101 (2015).

\bibitem{Tannenbaum1956}
E.~Tannenbaum, R.~J. Myers, and W.~D. Gwinn.
\newblock {Microwave Spectra, Dipole Moment, and Barrier to Internal Rotation
  of CH$_3$NO$_2$ and CD$_3$NO$_2$}.
\newblock \emph{J. Chem. Phys.} \textbf{25}, 42 (1956).

\bibitem{Jones1968}
W.~J. Jones and N.~Sheppard.
\newblock {The Gas-Phase Infrared Spectra of Nitromethane and Methyl Boron
  Difluoride; Fine Structure Caused by Internal Rotation}.
\newblock \emph{Proc. R. Soc. A Math. Phys. Eng. Sci.} \textbf{304}, 135--155
  (1968).

\bibitem{Cox1972}
A.~P. Cox and S.~Waring.
\newblock {Microwave spectrum and structure of nitromethane}.
\newblock \emph{J. Chem. Soc. Faraday Trans. 2} \textbf{68}, 1060 (1972).

\bibitem{Rohart1975}
F.~Rohart.
\newblock {Microwave spectrum of nitromethane internal rotation Hamiltonian in
  the low barrier case}.
\newblock \emph{J. Mol. Spectrosc.} \textbf{57}, 301--311 (1975).

\bibitem{Sorensen1983}
G.~S{\o}rensen, T.~Pedersen, H.~Dreizler, A.~Guarnieri, and A.~P. Cox.
\newblock {Microwave spectra of nitromethane and D3-nitromethane}.
\newblock \emph{J. Mol. Struct.} \textbf{97}, 77--82 (1983).

\bibitem{Sorensen1983a}
G.~S{\o}rensen and T.~Pedersen.
\newblock {Symmetry and microwave spectrum of nitromethane}.
\newblock \emph{Stud. Phys. Theor. Chem.} \textbf{23}, 219--236 (1983).

\bibitem{Light2000}
J.~C. Light and T.~{Carrington Jr.}
\newblock {Discrete-Variable Representations and their Utilization}.
\newblock \emph{Adv. Chem. Phys.} \textbf{114}, 263 (2000).

\bibitem{Christoffel1982}
K.~M. Christoffel and J.~M. Bowman.
\newblock {Investigations of self-consistent field, SCF CI and virtual state
  configuration interaction vibrational energies for a model three-mode
  system}.
\newblock \emph{Chem. Phys. Lett.} \textbf{85}, 220--224 (1982).

\bibitem{Carter1998a}
S.~Carter, J.~M. Bowman, and N.~C. Handy.
\newblock {Extensions and tests of ``multimode'': a code to obtain accurate
  vibration/rotation energies of many--mode molecules}.
\newblock \emph{Theor. Chem. Accounts} \textbf{100}, 191--198 (1998).

\bibitem{VanVleck1929}
J.~H. {Van Vleck}.
\newblock {On $\sigma$-Type Doubling and Electron Spin in the Spectra of
  Diatomic Molecules}.
\newblock \emph{Phys. Rev.} \textbf{33}, 467--506 (1929).

\bibitem{Wilson1936}
E.~B. Wilson and J.~B. Howard.
\newblock {The Vibration-Rotation Energy Levels of Polyatomic Molecules I.
  Mathematical Theory of Semirigid Asymmetrical Top Molecules}.
\newblock \emph{J. Chem. Phys.} \textbf{4}, 260 (1936).

\bibitem{Nielsen1951}
H.~H. Nielsen.
\newblock {The Vibration-Rotation Energies of Molecules}.
\newblock \emph{Rev. Mod. Phys.} \textbf{23}, 90--136 (1951).

\bibitem{Watson1967}
J.~K.~G. Watson.
\newblock {Determination of Centrifugal Distortion Coefficients of
  Asymmetric-Top Molecules}.
\newblock \emph{J. Chem. Phys.} \textbf{46}, 1935 (1967).

\bibitem{Watson1977}
J.~K.~G. Watson.
\newblock {Aspects of quartic and sextic centrifugal effects on rotational
  energy levels}.
\newblock In \emph{Vib. Spectra Struct. Vol. 6}, edited by J.~Durig, chapter~1
  (Elsevier, Amsterdam, 1977).

\bibitem{Field2011}
R.~W. Field, J.~H. Baraban, S.~H. Lipoff, and A.~R. Beck.
\newblock {Effective Hamiltonians for Electronic Fine Structure and Polyatomic
  Vibrations}.
\newblock In \emph{Handb. High-Resolution Spectrosc.}, edited by M.~Quack and
  F.~Merkt, page 1461 (John Wiley \& Sons, Chichester, UK, 2011).

\bibitem{Gordy1984}
W.~Gordy and R.~L. Cook.
\newblock \emph{{Microwave Molecular Spectra}} (John Wiley \& Sons, New York,
  1984), 3rd edition.

\bibitem{Podolsky1928}
B.~Podolsky.
\newblock {Quantum-Mechanically Correct Form of Hamiltonian Function for
  Conservative Systems}.
\newblock \emph{Phys. Rev.} \textbf{32}, 812--816 (1928).

\bibitem{Lauvergnat2002}
D.~Lauvergnat and A.~Nauts.
\newblock {Exact numerical computation of a kinetic energy operator in
  curvilinear coordinates}.
\newblock \emph{J. Chem. Phys.} \textbf{116}, 8560 (2002).

\bibitem{Fabri2011}
C.~F{\'{a}}bri, E.~M{\'{a}}tyus, and A.~G. Cs{\'{a}}sz{\'{a}}r.
\newblock {Rotating full- and reduced-dimensional quantum chemical models of
  molecules.}
\newblock \emph{J. Chem. Phys.} \textbf{134}, 074105 (2011).

\bibitem{Lukka1995}
T.~J. Lukka.
\newblock {A simple method for the derivation of exact quantum-mechanical
  vibration-rotation Hamiltonians in terms of internal coordinates}.
\newblock \emph{J. Chem. Phys.} \textbf{102}, 3945--3955 (1995).

\bibitem{Watson2004}
J.~K.~G. Watson.
\newblock {The molecular vibration-rotation kinetic-energy operator for general
  internal coordinates}.
\newblock \emph{J. Mol. Spectrosc.} \textbf{228}, 645--658 (2004).

\bibitem{Wilson1955}
E.~B. {Wilson Jr.}, J.~C. Decius, and P.~C. Cross.
\newblock \emph{{Molecular Vibrations}} (Dover, New York, 1980).

\bibitem{Dymarsky2005}
A.~Y. Dymarsky and K.~N. Kudin.
\newblock {Computation of the pseudorotation matrix to satisfy the Eckart axis
  conditions}.
\newblock \emph{J. Chem. Phys.} \textbf{122}, 124103 (2005).

\bibitem{Szalay2015}
V.~Szalay.
\newblock {Aspects of the Eckart frame ro-vibrational kinetic energy operator}.
\newblock \emph{J. Chem. Phys.} \textbf{143}, 064104 (2015).

\bibitem{Szalay2015a}
V.~Szalay.
\newblock {Understanding nuclear motions in molecules: Derivation of Eckart
  frame ro-vibrational Hamiltonian operators via a gateway Hamiltonian
  operator.}
\newblock \emph{J. Chem. Phys.} \textbf{142}, 174107 (2015).

\bibitem{Lauvergnat2016}
D.~Lauvergnat, J.~M. Luis, B.~Kirtman, H.~Reis, and A.~Nauts.
\newblock {Numerical and exact kinetic energy operator using Eckart conditions
  with one or several reference geometries: Application to HONO.}
\newblock \emph{J. Chem. Phys.} \textbf{144}, 084116 (2016).

\bibitem{Krasnoshchekov2014}
S.~V. Krasnoshchekov, E.~V. Isayeva, and N.~F. Stepanov.
\newblock {Determination of the Eckart molecule-fixed frame by use of the
  apparatus of quaternion algebra}.
\newblock \emph{J. Chem. Phys.} \textbf{140}, 154104 (2014).

\bibitem{NITROGEN}
{{NITROGEN, Numerical and Iterative Techniques for Rovibronic Energies with
  General Internal Coordinates, a program by P. B. Changala,
  http://www.colorado.edu/nitrogen}}.

\bibitem{CFOUR}
CFOUR, Coupled-cluster techniques for Computational Chemistry, a
  quantum-chemical program package by J.F. Stanton, J. Gauss, M.E. Harding,
  P.G. Szalay with contributions from A.A. Auer, R.J. Bartlett, U. Benedikt, C.
  Berger, D.E. Bernholdt, Y.J. Bomble, L. Cheng, O. Christiansen, M. Heckert,
  O. Heun, C. Huber, T.-C. Jagau, D. Jonsson, J. Jus\'{e}lius, K. Klein, W.J.
  Lauderdale, F. Lipparini, D.A. Matthews, T. Metzroth, L.A. M\"{u}ck, D.P.
  O'Neill, D.R. Price, E. Prochnow, C. Puzzarini, K. Ruud, F. Schiffmann, W.
  Schwalbach, C. Simmons, S. Stopkowicz, A. Tajti, J. V\'{a}zquez, F. Wang,
  J.D. Watts and the integral packages MOLECULE (J. Alml\"{o}f and P.R.
  Taylor), PROPS (P.R. Taylor), ABACUS (T. Helgaker, H.J. Aa. Jensen, P.
  J{\o}rgensen, and J. Olsen), and ECP routines by A. V. Mitin and C. van
  W\"{u}llen. For the current version, see http://www.cfour.de.

\bibitem{Cernicharo2015}
J.~Cernicharo, M.~C. McCarthy, C.~A. Gottlieb, M.~Ag{\'{u}}ndez, L.~V. Prieto,
  J.~H. Baraban, P.~B. Changala, M.~Gu{\'{e}}lin, C.~Kahane, M.~A.
  Martin-Drumel, N.~A. Patel, N.~J. Reilly, J.~F. Stanton, G.~Quintana-Lacaci,
  S.~Thorwirth, and K.~H. Young.
\newblock {Discovery of SiCSi in IRC+10216: A Missing Link Between Gas and Dust
  Carriers of Si--C Bonds}.
\newblock \emph{Astrophys. J. Lett.} \textbf{806}, L3 (2015).

\bibitem{McCarthyComm}
M. C. McCarthy, private communication (2016).

\bibitem{Puzzarini2008}
C.~Puzzarini, M.~Heckert, and J.~Gauss.
\newblock {The accuracy of rotational constants predicted by high-level
  quantum-chemical calculations. I. molecules containing first-row atoms}.
\newblock \emph{J. Chem. Phys.} \textbf{128}, 194108 (2008).

\bibitem{Wang2015}
X.~Wang, S.~Carter, and J.~M. Bowman.
\newblock {Pruning the Hamiltonian Matrix in MULTIMODE: Test for C$_2$H$_4$ and
  Application to CH$_3$NO$_2$ Using a New Ab Initio Potential Energy Surface.}
\newblock \emph{J. Phys. Chem. A} \textbf{119}, 11632--11640 (2015).

\bibitem{Almlof1987a}
J.~Almlöf and P.~R. Taylor.
\newblock {General contraction of Gaussian basis sets. I. Atomic natural
  orbitals for first- and second-row atoms}.
\newblock \emph{J. Chem. Phys.} \textbf{86}, 4070 (1987).

\bibitem{McCaslin2013}
L.~McCaslin and J.~Stanton.
\newblock {Calculation of fundamental frequencies for small polyatomic
  molecules: a comparison between correlation consistent and atomic natural
  orbital basis sets}.
\newblock \emph{Mol. Phys.} \textbf{111}, 1492--1496 (2013).

\bibitem{Dawadi2015}
M.~B. Dawadi, S.~Twagirayezu, D.~S. Perry, and B.~E. Billinghurst.
\newblock {{High-resolution Fourier transform infrared synchrotron spectroscopy
  of the NO$_2$ in-plane rock band of nitromethane}}.
\newblock \emph{J. Mol. Spectrosc.} \textbf{315}, 10--15 (2015).

\bibitem{Spaun2016}
B.~Spaun, P.~B. Changala, D.~Patterson, B.~J. Bjork, O.~H. Heckl, J.~M. Doyle,
  and J.~Ye.
\newblock {Continuous probing of cold complex molecules with infrared frequency
  comb spectroscopy}.
\newblock \emph{Nature} \textbf{533}, 517--520 (2016).

\end{thebibliography}


\appendix*
\section{}
During the procedure described in Section~\ref{sec:eckart} for determining the (quasi-)Eckart frame KEO, we required the derivatives of $\vec{\gamma}$, the eigenvector of the $4\times4$ matrix $C$. In this appendix, we derive the analytical, closed form expressions for these derivatives. To more formally state the problem, we begin, using ket notation, with a non-degenerate eigenvector $| n \rangle$ and its eigenvalue $\lambda_n$ of a real symmetric matrix $H$, which are related by
\begin{equation}
H | n \rangle = \lambda_n | n \rangle \label{eq:diffHn},
\end{equation}
where $H$, $| n \rangle$, and $\lambda_n$ are all functions of a set of real parameters $\vec{p} =( p^1, p^2, p^3, \ldots)$,
\begin{equation}
H(\vec{p}) | n(\vec{p}) \rangle = \lambda_n(\vec{p}) | n (\vec{p})\rangle.
\end{equation}
We are interested in computing the first and second derivatives of $| n \rangle$ and $\lambda_n$ with respect to these parameters given knowledge of the derivatives of $H$. The final results are given by Eqs.~\ref{eq:dlambda}, \ref{eq:dn}, \ref{eq:ddlambda}, and \ref{eq:ddn}. 

\subsection{Normalization relations}
First, we derive some results based on the condition that $\langle n | n \rangle = 1$ for all values of $\vec{p}$. We begin by expanding $| n \rangle$ as a function of $\vec{p}$ about some reference point $\vec{p}_0$. Defining $\delta \vec{p} = \vec{p} - \vec{p}_0$, we have
\begin{align}
| n \rangle = | n(\vec{p}) \rangle = | n \rangle_0 + \delta p^i | \partial_i n \rangle_0 + \frac{1}{2}\delta p^i\delta p^j | \partial_i \partial_j n \rangle_0 + \ldots,
\end{align}
where the subscript 0 indicates evaluation at $p^i = p^i_0$ and summation over repeated indices is implied. We now expand the inner product $\langle n | n \rangle = 1$ similarly:
\begin{align}
\Big(\langle n |_0 + \delta p^i \langle \partial_i n |_0 + \frac{1}{2}\delta p^i\delta p^j \langle \partial_i \partial_j n |_0 + \ldots\Big) &\times \nonumber\\
\Big(| n \rangle_0 + \delta p^i | \partial_i n \rangle_0 + \frac{1}{2}\delta p^i\delta p^j | \partial_i \partial_j n \rangle_0 &+ \ldots\Big) = 1.
\end{align}
Collecting terms of the same order in $\delta p^i$ yields
\begin{align}
\langle n | n \rangle_0 &= 1\\
\delta p^i \langle \partial_i n | n \rangle_0 + \delta p^i \langle n | \partial_i n \rangle_0 &= 0\\
\frac{1}{2}\delta p^i\delta p^j \langle \partial_i\partial_j n| n \rangle_0  + \frac{1}{2}\delta p^i\delta p^j \langle n | \partial_i\partial_j n \rangle_0&\nonumber\\
\qquad+\delta  p^i \delta p^j \langle \partial_i n | \partial_j n\rangle_0 &= 0.
\end{align}

Noting that (i) these conditions are true for an arbitrary reference point $\vec{p}_0$ and displacements $\delta p^i$, and (ii) that $H$ is real, so that we can choose the eigenvectors to be real and therefore $\langle m | n \rangle$ = $\langle n | m \rangle$, we have the following two identities regarding the projection of the derivatives of an eigenvector on itself.
\begin{align}
\langle n | \partial_i n \rangle &= 0 \label{eq:ndn}\\
\langle n | \partial_i \partial_j n \rangle &= - \langle \partial_i n | \partial_j n \rangle \label{eq:nddn}
\end{align}

\subsection{First derivatives}
Taking the derivative of \ref{eq:diffHn} with respect to parameter $p^i$ yields
\begin{align}
\partial_i H | n \rangle + H | \partial_i n\rangle &= \partial_i \lambda_n | n \rangle + \lambda_n | \partial_i n\rangle.
\end{align}
After taking the inner product with $\langle n |$, we have
\begin{align}
\langle n | \partial_i H | n \rangle + \lambda_n\langle n  | \partial_i n\rangle &= \partial_i \lambda_n \langle n | n \rangle + \lambda_n \langle n | \partial_i n\rangle \nonumber \\
\Rightarrow \partial_i \lambda_n &= \langle n | \partial_i H | n \rangle\label{eq:dlambda}.
\end{align}
If instead, we take the inner product with a different eigenvector $\langle m | \neq \langle n |$ (with $\lambda_m \neq \lambda_n$), we have
\begin{align}
\langle m | \partial_i H | n \rangle + \langle m | H | \partial_i n \rangle &= \partial_i \lambda_n \langle m | n \rangle + \lambda_n \langle m | \partial_i n \rangle\nonumber \\
\langle m | \partial_i H | n \rangle + \lambda_m \langle m | \partial_i n \rangle &= 0 + \lambda_n \langle m | \partial_i n \rangle \nonumber \\
\Rightarrow \langle m | \partial_i n \rangle &= \frac{\langle m | \partial_i H | n \rangle}{\lambda_n - \lambda_m} \label{eq:mdn}.
\end{align}
This relation and Eq.~\ref{eq:ndn} together completely specify $| \partial_i n \rangle$:
\begin{align}
| \partial_i n \rangle &= \sum_{m\neq n} \frac{\langle m | \partial_i H | n \rangle}{\lambda_n - \lambda_m} | m \rangle \label{eq:dn}.
\end{align}

\subsection{Second derivatives}
Taking the mixed second derivatives of \ref{eq:diffHn} and multiplying by $\langle n |$ gives
\begin{align}
&\langle n | \partial_i \partial_j H | n \rangle + \langle n | \partial_i H | \partial_j n \rangle + \hphantom{a}\nonumber\\
&\qquad\langle n | \partial_j H | \partial_i n \rangle + \langle n | H | \partial_i \partial_j n \rangle =\nonumber\\
&\partial_i \partial_j \lambda_n \langle n | n \rangle + \partial_i \lambda_n \langle n | \partial_j n \rangle  + \hphantom{a}\nonumber \\
&\qquad\partial_j \lambda_n \langle n | \partial_i n \rangle + \lambda_n \langle n | \partial_i \partial_j n \rangle
\end{align}
\begin{align}
&\rightarrow \partial_i \partial_j \lambda_n = \langle n | \partial_i \partial_j H | n \rangle +\hphantom{a}\nonumber\\
& \qquad  \langle n | \partial_i H | \partial_j n \rangle + \langle n | \partial_j H | \partial_i n \rangle.
\end{align}
We now insert the resolution of the identity into the last two terms,
\begin{align}
\partial_i \partial_j \lambda_n &= \langle n | \partial_i \partial_j H | n \rangle + \hphantom{a}\nonumber\\
& \sum_m \big[ \langle n | \partial_i H | m \rangle \langle m | \partial_j n \rangle + \langle n | \partial_j H|  m\rangle \langle m | \partial_i n \rangle \big].
\end{align}
The terms in the sum with $m=n$ are equal to zero by Eq.~\ref{eq:ndn}. Those with $m\neq  n$ can be evaluated with Eq.~\ref{eq:mdn}. This results in
\begin{align}
\partial_i \partial_j \lambda_n &= \langle n | \partial_i \partial_j H | n \rangle + \hphantom{a}\nonumber\\
& 2 \sum_{m\neq n} \left[ \frac{\langle n | \partial_i H | m \rangle \langle m | \partial_j H | n \rangle}{\lambda_n - \lambda_m} \right]\label{eq:ddlambda}.
\end{align}

We repeat this again multiplying with $\langle m | \neq \langle n | $ instead:
\begin{align}
&\langle m | \partial_i \partial_j H | n \rangle + \langle m | \partial_i H | \partial_j n \rangle + \hphantom{a}\nonumber\\
&\qquad \langle m | \partial_j H | \partial_i n \rangle + \langle m | H | \partial_i \partial_j n \rangle =\nonumber\\
&\partial_i \partial_j \lambda_n \langle m | n \rangle + \partial_i \lambda_n \langle m | \partial_j n \rangle + \hphantom{a}\nonumber\\
&\qquad \partial_j \lambda_n \langle m | \partial_i n \rangle + \lambda_n \langle m | \partial_i \partial_j n \rangle.
\end{align}
To continue, we insert the resolution of the identity into the second and third terms of the left-hand side and use Eq.~\ref{eq:mdn}. We also use Eqs.~\ref{eq:dlambda} and \ref{eq:mdn} in the second and third terms of the right-hand side to obtain
\begin{widetext}
\begin{align}
\langle m | \partial_i \partial_j H | n \rangle + \sum_{\ell \neq n}\left[\frac{\langle m| \partial_i H | \ell \rangle \langle \ell | \partial_j H | n \rangle}{\lambda_n - \lambda_\ell} + \frac{\langle m| \partial_j H | \ell \rangle \langle \ell | \partial_i H | n \rangle}{\lambda_n - \lambda_\ell} \right] + \lambda_m \langle m | \partial_i \partial_j n \rangle =\nonumber\\
0+ \frac{\langle n | \partial_i H | n \rangle \langle m | \partial_j H | n \rangle}{\lambda_n - \lambda_m}+ \frac{\langle n | \partial_j H | n \rangle \langle m | \partial_i H | n \rangle}{\lambda_n - \lambda_m} + \lambda_n \langle m | \partial_i \partial_j n \rangle.
\end{align}
Assuming again that $\lambda_m \neq \lambda_n$, we arrive at
\begin{align}
\langle m | \partial_i \partial_j n \rangle &= \frac{\langle m | \partial_i \partial_j H | n \rangle}{\lambda_n - \lambda_m} + \sum_{\ell \neq n}\left[\frac{\langle m| \partial_i H | \ell \rangle \langle \ell | \partial_j H | n \rangle + \langle m| \partial_j H | \ell \rangle \langle \ell | \partial_i H | n \rangle}{(\lambda_n - \lambda_\ell)(\lambda_n - \lambda_m)} \right] \nonumber\\
&\qquad - \frac{\langle n | \partial_i H | n \rangle \langle m | \partial_j H | n \rangle + \langle n | \partial_j H | n \rangle \langle m | \partial_i H | n \rangle}{(\lambda_n - \lambda_m)^2}.
\end{align}
\end{widetext}
This expression gives us the projection of $| \partial_i \partial_j n \rangle$ onto eigenvectors $| m \rangle \neq | n \rangle$. Unlike the case with the first derivative, there is a non-zero projection onto $| n \rangle$ itself, given by Eq.~\ref{eq:nddn}. Inserting another resolution of the identity into that expression gives
\begin{align}
\langle n | \partial_i \partial_j n \rangle &= - \sum_{\ell} \langle \partial_i n | \ell \rangle \langle \ell | \partial_j n \rangle\nonumber\\
&= - \sum_{\ell\neq n } \frac{\langle n | \partial_i H | \ell \rangle \langle \ell | \partial_j H | n \rangle}{(\lambda_n-\lambda_\ell)^2}.
\end{align}

We now have all the information needed to construct the second derivatives of the eigenvectors. Our final result is
\begin{align}
&| \partial_i \partial_j n \rangle = \sum_{m\neq n} \left\{ \frac{\langle m | \partial_i \partial_j H | n \rangle}{\lambda_n - \lambda_m}\right.\nonumber\\
&+ \sum_{\ell \neq n}\left[\frac{\langle m| \partial_i H | \ell \rangle \langle \ell | \partial_j H | n \rangle + \langle m| \partial_j H | \ell \rangle \langle \ell | \partial_i H | n \rangle}{(\lambda_n - \lambda_\ell)(\lambda_n - \lambda_m)} \right]\nonumber\\
&\left. - \frac{\langle n | \partial_i H | n \rangle \langle m | \partial_j H | n \rangle + \langle n | \partial_j H | n \rangle \langle m | \partial_i H | n \rangle}{(\lambda_n - \lambda_m)^2}\right\}|m\rangle\nonumber\\
& - \left( \sum_{\ell\neq n } \frac{\langle n | \partial_i H | \ell \rangle \langle \ell | \partial_j H | n \rangle}{(\lambda_n-\lambda_\ell)^2}\right) | n \rangle \label{eq:ddn}.
\end{align}
\end{document}